# GETTING CRITICAL: MAKING SENSE OF THE EU CYBERSECURITY FRAMEWORK FOR CLOUD PROVIDERS[+]

*Ian Walden*[*] *and Johan David Michels*[**]

**ABSTRACT**

*In this chapter, we review how the EU cybersecurity regulatory framework impacts providers of cloud computing services. We examine the evolving regulatory treatment of cloud services as an enabler of the EU's digital economy and question whether all cloud services should be treated as critical infrastructure. Further, we look at how the safeguarding and incident notification obligations under the General Data Protection Regulation ('GDPR') and the Network and Information Systems Directive ('NISD') apply to cloud providers.*

[+] This paper is an author-edited, pre-print version of a book chapter that will appear in: Andrader F, Abreu J, and Freitas P, *Legal Developments in Cyber-Security and Related Fields*, (Forthcoming, Springer). The chapter has been produced by members of the Cloud Legal Project, Centre for Commercial Law Studies, Queen Mary University of London. The authors are grateful to Microsoft for the generous financial support that has made this project possible. Responsibility for views expressed, however, remains with the authors.

[*] Professor of Information and Communications Law and Director, Centre for Commercial Law Studies, Queen Mary University of London.

[**] Researcher and PhD Candidate, Cloud Legal Project, Centre for Commercial Law Studies, Queen Mary University of London.

*We also consider the proposed revision of the NISD and look at newly developed voluntary assurance mechanisms for cloud providers, including codes of conduct and certification schemes. We conclude that, since cloud providers are typically subject to both NISD and GDPR and to the jurisdiction of multiple regulators, they face divergent regulatory approaches, which can lead to unintended outcomes and high compliance costs.*

*March 2022*

**TABLE OF CONTENTS**

# 1. INTRODUCTION

The EU cybersecurity regulatory framework has developed rapidly over recent years, both in terms of breadth and depth. In 2018, the General Data Protection Regulation ('GDPR')[1] and the Network and Information Systems Directive ('NISD')[2] came into force. In 2019, the EU Cybersecurity Act followed.[3] Currently, discussions are ongoing around the Commission's proposals for a revised NIS2 Directive ('draft NIS2').[4] In this chapter, we review what these developments mean for providers of cloud computing services. We start by examining the evolving treatment of cloud services as an enabler of the EU's digital economy and as critical infrastructure. Then, we review the two core components of the EU cybersecurity framework, namely: (i) safeguarding obligations, including voluntary schemes of assurance, and (ii) incident notification obligations. Finally, we consider issues of jurisdiction, oversight, and enforcement. Throughout, we reflect critically on the evolving regulatory framework and highlight strengths and shortcomings.

Our focus is on the security requirements the GPDR and NISD impose on cloud *providers* directly. We do not cover EU legislation that imposes security requirements on cloud *customers* and so might impact cloud providers indirectly, for instance through contract negotiations. For example, customers in regulated sectors, such as financial service providers, typically need to comply with security requirements throughout their IT supply chain.[5] Finally, we do not cover the proposed Directive on Critical Entities' Resilience ('CERD'), which will replace the European Critical Infrastructure Directive.[6] The CERD is not expected to impose new obligations on cloud providers, whose physical security is already covered under the NISD.[7] Other elements of the EU cybersecurity framework beyond the scope of this chapter include strengthening and harmonizing criminal law,[8] and the provision of trust services.[9]

---

[1] Regulation (EU) 2016/679 on the protection of natural persons with regard to the processing of personal data and on the free movement of such data (2016), OJ L 119 4.5.2016 ('GDPR').

[2] Directive (EU) 2016/1148 concerning measures for a high common level of security of network and information systems across the Union [2016], OJ L 194/1 19.7.2016 ('NISD').

[3] Regulation (EU) 2019/881 on ENISA (the European Union Agency for Cybersecurity) and on information and communications technology cybersecurity certification and repealing Regulation (EU) No 526/2013, OJ L 151 7.6.2019 ('Cybersecurity Act').

[4] EU Commission, 'Proposal for a Directive on measures for a high common level of cybersecurity across the Union, repealing Directive (EU) 2016/1148', 16 December 2020, COM(2020) 823 final ('NIS 2 Proposal').

## 2. SCOPE: CLOUD AS A REGULATED ACTIVITY

In a little over 10 years, cloud computing has become the dominant paradigm for data processing. Survey evidence indicates that, as of 2020, 90% of companies used cloud services, running half of their workload and storing half of their data in the public cloud.[10] As a result, the market for public cloud services grew to $270bn in 2020.[11] As cloud computing spread, it raised challenges for applying existing laws and regulations, not least with respect to data protection laws.[12] However, until recently, cloud computing was not a distinct regulated activity, separate from other models for processing data, such as outsourcing. That changed in 2018, as the NISD subjected the provision of cloud services to separate regulatory treatment in the context of cybersecurity.

---

[5] EU Commission, 'Proposal for a Regulation on 'digital operational resilience for the financial sector and amending Regulations (EC) No 1060/2009, (EU) No 648/2012, (EU) No 600/2014 and (EU) No 909/2014', 24 September 2020, COM(2020) 595 final.

[6] Directive 2008/114/EC on the identification and designation of European critical infrastructures and the assessment of the need to improve their protection [2008] OJ L 345 23.12.2008.

[7] EU Commission, 'Proposal for a Directive on the resilience of critical entities', 16 December 2020, COM(2020) 829 final, p.3; Recital 14 and Art. 7, Proposed Directive on the resilience of critical entities.

[8] E.g. Directive 2013/40/EU on attacks against information systems, L 218/8, 14.8.2013.

[9] Regulation (EU) 2014/910 on electronic identification and trust services for electronic transactions in the internal market, L 257/73, 28.8.2014.

Cloud computing offers data processing capacity in a flexible, efficient, and readily-accessible manner.[13] Accessing remote computing resources depends on connectivity over communications networks such as the public internet. As examined below, this communications network infrastructure is treated as a separate regulated activity from the networking that comprises part of a cloud computing service. Cloud computing is typically distinguished into three service types: infrastructure-as-a-service ('IaaS'), platform-as-a-service ('PaaS') and software-as-a-service ('SaaS').[14] The deployment of cloud computing can also be divided into different models, such as public, private, and community cloud, as well as hybrid implementations.[15] While these distinctions do not fully reflect the reality of this large and diverse sector, they aid legal and policy analysis by highlighting different technical and market conditions.

---

[10] Flexera, 'State of the Cloud Report 2020', (2020) https://info.flexera.com/SLO-CM-REPORT-State-of-the-Cloud-2020; Ponemon Institute, 'Protecting Data in the Cloud' (2019) https://safenet.gemalto.com/cloud-security-research/.

[11] Gartner, 'Gartner Forecasts Worldwide Public Cloud End-User Spending to Grow 23% in 2021', 21 April 2021, https://www.gartner.com/en/newsroom/press-releases/2021-04-21-gartner-forecasts-worldwide-public-cloud-end-user-spending-to-grow-23-percent-in-2021.

[12] See generally Chapters 8-11 in Millard C (ed.) *Cloud Computing Law* (2021, 2nd edn., Oxford University Press).

[13] National Institute of Standards and Technology, U.S. Department of Commerce, 'Evaluation of Cloud Computing Services Based on NIST SP 800-145, Special Publication 500-322' (2018) ('NIST Publication 500-322').

[14] See, for example, ISO/IEC 17788:2014 'Cloud Computing – Overview and Vocabulary'.

The NISD imposes obligations specifically on providers of cloud computing services, while the GDPR is of general application, but creates complex issues when applied to some cloud providers. This section critically examines this twin-track regulatory approach. Given the prevalence of cloud computing, we question whether it is useful to distinguish between the regulation of cloud services specifically on the one hand, and generic behaviours, such as processing personal data, on the other. This approach can create a regulatory mess that obscures rather than clarifies cybersecurity obligations and imposes unnecessary additional compliance burdens.

## 2.1 CLOUD PROVIDERS AS REGULATED PROVIDERS OF CRITICAL INFRASTRUCTURE

Under the NISD, the term 'networks' is shorthand for electronic communication networks,[16] as defined under the New Regulatory Framework,[17] although public electronic communications networks (*read:* external service providers) are excluded from the NISD regime.[18] By contrast, the notion of 'information systems' encompasses both on-premises devices for 'processing digital data', as well as the external provision of processing capacity in the form of services such as cloud computing.[19] Cloud computing services primarily fall in the category of information society services (or 'ISSs'), regulated under the eCommerce Directive.[20] This covers services provided by electronic means, at a distance, and at the request of the service recipient. These services are also referred to as 'OTT', since they are provided 'over-the-top' of the underlying electronic communications networks.

---

[15] NIST, 'Publication 500-322'.

[16] Art. 4(1)(a) NISD.

[17] Directive 2002/21/EC on a common regulatory framework for electronic communications networks and services, OJ L 108/33, 24.4.2002 ('Framework Directive'), as amended.

[18] Art. 1(3) NISD.

Telecoms networks typically had monopolistic origins and a utility-like nature. As a result, security (in terms of confidentiality, integrity and availability) has always been a key element of the EU regulatory framework. It was traditionally described as an 'essential requirement', embodying the 'general public interest'.21 However, until 2009, the regulator was tasked with ensuring security,22 through instruments such as prior authorisation.23 This changed with the 2009 reform, which imposed safeguarding and incident notification obligations directly on operators.24 In addition, sectoral data protection rules imposed another layer of security

---

[19] Art. 4(1)(b) NISD. The definition is taken from Directive 2013/40/EU 'on attacks against information systems' OJ L 218/8/, 14.8.2013.

[20] Directive 2000/31/EC on certain legal aspects of information society services, in particular electronic commerce, in the Internal Market, OJ L 178/1, 17.7.2000.

[21] See Walden I 'European Union Communications Law', in: *Telecommunications Law and Regulation* (2018, 5th edn., Oxford University Press), at 4.4.2.

[22] Art. 8(4)(f) Framework Directive: "ensuring that the integrity and security of public communications networks are maintained".

[23] Directive 2002/20/EC on the authorization of electronic communications networks and services OJ L 108/21, 24.4.2002, Annex, A.16: "Security of public networks against unauthorised access".

[24] The Framework Directive was amended in 2009 by Directive 2009/140/EC, which inserted Chapter IIIa: Security and Integrity of Networks and Services.

obligations on providers of 'publicly available electronic communication services'.25

In contrast, the regulation of ISSs started from the opposite end of the spectrum. In the early 2000s, these newly emerging services were subjected to a light-touch regulatory approach, designed primarily to facilitate market development. At the height of the first dot.com boom, security was not considered a key regulatory need, because the policy imperatives for ISSs were early adoption, diversity, innovation, and consumer choice.

The regulation of ISSs changed with the NISD and the reform of telecommunications law. First, the NISD imposed safeguarding and notification obligations on digital services providers (or 'DSPs'), which included cloud providers. Second, the European Electronic Communications Code extended the scope of 'electronic communications services', and the accompanying safeguarding and notification obligations, to encompass operators of 'interpersonal communication services', which includes cloud-based messaging services.26 These developments represented a fundamental regulatory shift for cloud providers, aligning their regulatory obligations more closely with those of telecoms network operators. The marketplace of ISSs has also fundamentally changed over the past two decades, with increased uptake of cloud services, as well as the development of complex supply chains and ever more nuanced service differentiation.

This raises the question: should all cloud services be treated as akin to critical infrastructure? The Commission's

---

[25] Art. 4 Directive 02/58/EC concerning the processing of personal data and the protection of privacy in the electronic communications sector, OJ L 201/37, 31.7.2002, as amended.

[26] Chapter V Directive (EU) 2018/1972 establishing the European Electronic Communications Code, OJ L 321/36, 17.12.2018.

original 2013 proposal for an NISD had listed six types of ISSs:

1. e-commerce platforms;
2. internet payment gateways;
3. social networks;
4. search engines;
5. cloud computing services; and,
6. application stores.[27]

This list was intended to be non-exhaustive, meaning that the measures could extend to other types of services.[28] According to the Commission, the distinguishing criterion was ISSs that "enable the provision of other information services", meaning services which "underpin downstream information society services or on-line activities".[29] This would suggest a focus on business-to-business ('B2B') services, rather than business-to-consumer ('B2C') services. The proposal further explicitly excluded "software developers and hardware manufacturers", a distinction that remains in the final version.[30] In the final version of the NISD, the list of DSPs became exhaustive and was trimmed down from six to three. Yet the NISD rationale for the inclusion of the listed categories of DSPs became even more vague. The limitation to supply chains for the provision of other ISSs disappeared, since all types of persons and entities rely on DSP services.

---

[27] EU Commission, 'Proposal for a Directive concerning measures to ensure a high common level of network and information security across the Union' COM(2013) 48 final, 7.2.2013 ('Proposal'), at Annex II.

[28] Art. 3(8)(a) Proposal.

[29] Recital 24 Proposal.

[30] Recital 50 NISD.

Instead, the NISD stated that such services "*could be* an important resource for their users" and therefore reliance may be a source of vulnerability as "users *might not always* have alternatives available" [emphasis added]. Further, "many businesses in the Union increasingly rely" on DSPs "for the smooth functioning" of their businesses. As a result, disruption of DSP services could impact "key economic and societal activities in the Union".[31]

At face value, this seems a remarkably weak basis upon which to justify the imposition of a regulatory regime on all cloud services, without evidence proffered to support it. The stated rationale suggests two possible underlying factors: (i) widespread uptake and reliance; and (ii) limited alternative suppliers. As noted above, survey evidence indicates that the use of cloud services is widespread. However, the degree to which users rely on such services "for the smooth functioning of their business" would seem to differ per service, depending on its functionality. SaaS services in particular support a wide range of different operational functions, from advertising on social media to backing-up files. Not all these functions are equally 'critical' to their customer's "smooth functioning".

In terms of limited available 'alternatives', this could be a result of market concentration and/or vendor lock-in, due to high switching costs. While a full market assessment lies beyond the scope of this paper, it could be argued that there is significant market concentration in the IaaS sector, particularly in the form of the hyperscale providers AWS, Microsoft Azure, and Google Cloud.[32] For example, these three hyperscale providers are estimated to hold around 60% of the IaaS market, while the top eight providers hold around 80% of the market.³³ Similar to telecoms networks, the IaaS market features high sunk costs and therefore might tend towards oligopoly. However, market conditions appear to differ at the PaaS and SaaS layers, which feature many smaller providers. Some commentators have expressed concerns that market power in the upstream IaaS market might enable hyperscale providers to influence the downstream markets for PaaS and SaaS.³⁴ Yet such concerns apply only to vertically integrated IaaS providers who are also active in PaaS and SaaS markets, not to SaaS providers active in downstream markets only. In sum, a competition-led rationale would seem difficult to justify given the current scale and diversity of the market, especially at the SaaS layer. What's more, imposing additional cybersecurity obligations could have the counterproductive effect of entrenching market concentration by increasing regulatory costs that are borne most easily by larger operators.³⁵

---

[31] Recital 48 NISD.

[32] These hyperscale IaaS providers do face increasing competition from Chinese providers such as Alibaba and Tencent.

Further, the extent of cloud vendor lock-in also appears to differ between cloud services, for two reasons. First, the ease of switching between cloud providers would depend on service type, as well as the applications deployed and the amount and format of data stored in the cloud. The on-demand, self-service nature of cloud services provides customers with flexibility that should support switching, although, in practice, migrating applications and large data sets can be technically challenging, costly and time-consuming. Nonetheless, regulation has been introduced to facilitate porting of data between service providers, designed to lower switching costs.[36] Second, the greater potential capability of cloud users to self-provision the computing resources received 'as-a-service' distinguishes cloud from other 'critical' infrastructures that are not generally replicable in-house. Finally, where a business completely relies on a single cloud provider for its "smooth functioning", such 'single-sourcing' would itself seem to be a poor security practice – and potentially an issue of corporate governance in terms of meeting obligations to ensure business continuity.[37] In sum, the argument for treating *all* cloud services as critical infrastructure, akin to telecoms networks, appears under-developed, at best, and weak, at worst.

---

³³ Statista, 'Digital Economy Compass 2021', p. 63 (based on research by Synergy Group). For a more detailed discussion, see Musin, T 'Estimation of Global Public IaaS market concentration by Linda index', (2021) SHS Web of Conferences 114, 01014 (2021), NTSSCEM 2021.

³⁴ See e.g. https://www.forbes.com/sites/peterbendor-samuel/2020/03/02/hyperscale-cloud-providers-shaping-the-platform-marketplace/?sh=7f031956103d.

³⁵ A similar argument can, and has, been made in respect of the GDPR.

Finally, the NISD distinction between software offered 'as-a-product' and 'as-a-service' also seems problematic, as well as raising an age-old debate in the field of computer law regarding liability for software faults.[38] The only justification proffered for excluding software developers and hardware manufacturers from the regime is that they are "already subject to existing rules on product liability".[39] Although liability incentives are one element of a comprehensive policy response to enhancing security, they are more complex and convoluted tools for signalling and facilitating behavioural change than the imposition of direct security obligations. This is even more true in respect of product liability, where the regime has largely fallen into disuse.[40] Further action is needed in this space and has begun, focused on specific types of equipment. In October 2021, the Commission published a proposal for a measure on the cybersecurity of radio equipment;[41] while in November 2021, the UK Government published a bill that would require manufacturers, importers or distributors of 'relevant connectable products', such as IoT devices, to implement certain security requirements.[42]

---

[36] With respect to measures designed to facilitate the porting of data between service providers, see Art. 6 Regulation (EU) 2018/1807 on a framework for the free flow of non-personal data in the European Union, OJ L303/59, 28.11.2018. See also the Art. 20 GDPR and Commission, Proposal for a Regulation on harmonised rules on fair access to and use of data (Data Act) COM(2022) 68 final 23.2.2022.

[37] See, for example, the US Securities and Exchange Commission, 17 CFR Parts 229 and 249 (Release Nos. 33-10459; 34-82746), 'Commission Statement and Guidance on Public Company Cybersecurity Disclosures' (18 February 2018). See also UK consultation on 'Restoring trust in audit and corporate governance' with regard to resilience statements (March 2021).

[38] See e.g. Samuelson P, 'Liability for Defective Electronic Information', (1993) Comm. of the ACM, p. 21; Pinkney K, 'Putting Blame Where Blame Is Due: Software Manufacturer and Customer Liability for Security-Related Software Failure', (2002) 13 ALB. L.J. SCI. & TECH, pp. 62-82.

The draft NIS2 would promote providers of cloud computing services to 'essential entities' (the new 'higher' category that replaces operators of essential services, or 'OES'), in the 'digital infrastructure' sector, together with data centre providers and content delivery network ('CDN') providers. In addition, the draft NIS2 removes the 'scale' requirement that previously applied to identifying and designating operators in relevant sectors as OES.[43] As a result, all cloud services would be treated as higher-category essential entities. Online marketplaces and online search engines remain 'important entities' (the new 'lower' category that replaces DSPs), together with social networks.[44] The recategorization is said to take into account both "the level of criticality of the sector or of the type of service, as well as the level of dependency of other sectors or types of service". Unfortunately, the Commission failed to provide analysis or evidence to indicate how these factors have been measured and assessed.[45] Treating all cloud services as essential (as opposed to merely important) is particularly problematic since, as argued above, the rationale for treating all cloud services as critical infrastructure is itself under-developed.

---

[39] Recital 48 NISD.

[40] Noto La Diega G and Walden I 'Contracting for the 'Internet of Things': Looking into the Nest', (2016) European Journal of Law and Technology, Vol. 7, No. 2.

[41] Commission Delegated Regulation supplementing Directive 2014/53/EU of the European Parliament and of the Council with regard to the application of the essential requirements referred to in Article 3(3), points (d), (e) and (f), of that Directive, C(2021) 7672 final, 29.10.2021.

[42] Product Security and Telecommunications Infrastructure Bill (24 November 2021), see further https://www.gov.uk/government/collections/the-product-security-and-telecommunications-infrastructure-psti-bill-factsheets.

Further, the draft NIS2 also alters the definition of cloud computing in three ways. First, it explicitly includes all deployment models, including private cloud. It is unclear whether the existing definition under NISD covers private cloud, which arguably does not involve access to an "elastic pool of shareable" resources (defined as different users being served "from the same underlying equipment"), since private cloud typically entails physically separated hardware for each customer.[46] We have previously argued that private cloud *should* be considered critical, since disruption of

---

[43] Art. 2 NIS 2. Under the NISD, OES were defined as entities in seven listed sectors that operated on such a scale that their service was "essential for the maintenance of critical societal and economic activities", depending on factors such as: the number of users and other sectors that depend on the service; its market share; the potential impact on economic and societal activities or public safety; and any alternative means for the provision of that service.

[44] Annexes, NIS 2 Proposal.

[45] Recital 11 NIS 2 Proposal.

[46] See Michels JD and Walden I 'Cybersecurity and Critical Infrastructure', in: Millard C (ed.) *Cloud Computing Law* (2021, 2nd edn., Oxford University Press), pp. 396-397.

private cloud could equally have far-reaching effects on key economic and societal activities, given the widespread adoption of corporate hybrid cloud strategies.[47] That said, the draft NIS2 definition retains the requirement of access to an "elastic pool of shareable" resources.[48] The definition therefore seems internally inconsistent. Second, cloud services must enable 'on-demand administration', which is described as the capability for a "user to unilaterally self-provision" resources.[49] While part of the standard NIST definition,[50] it is not immediately obvious why this characteristic is a necessary part of the new legal definition. The implication might be that such self-provision generates greater security risks, although it is unclear which cloud services would not qualify as "on-demand" and might therefore be excluded under the draft NIS2 definition. Third, cloud services must involve "remote access" to "distributed" resources. Distributed resources are defined as "located on different networked computers or devices and which communicate and coordinate among themselves by message passing".[51] Again, it is unclear why this characteristic is a necessary part of the new legal definition, although the additional requirement for "remote" access could be read as excluding on-premises private cloud.[52] Yet using cloud equipment located at the customer's premises could be viewed either as a security risk, or a mitigation measure. In sum the proposed amended definition of cloud introduces new ambiguities that might make it more

---

[47] Michels JD and Walden I (2021) n 46, p. 397.

[48] Recital 16 NIS 2 Proposal.

[49] Art. 4(19) NIS 2 Proposal.

[50] NIST, 'Publication 500-322', p. 3.

[51] Recital 16, NIS 2 Proposal.

[52] E.g. AWS Outposts.

difficult for service providers to determine whether they fall within the regulation's scope.[53]

## 2.2 CLOUD PROVIDERS AS REGULATED PROCESSORS OF PERSONAL DATA

The GDPR applies generally to the processing of personal data. Under the GDPR, security is a critical component of the compliance regime. Controllers have an obligation to comply with, and be able to demonstrate compliance with, the 'integrity and confidentiality' principle, by which they should ensure that any personal data is processed with 'appropriate security'.[54] This principle is then elaborated into specific obligations on both controller and processor to implement 'appropriate technical and organisational measures'; which are examined further below in Section 3. While controllers and processors share certain obligations, the compliance burden sits more heavily with the controller than the processor. It is therefore important to consider the different roles of cloud computing providers under the GDPR regime, as controller, joint controller, processor and/or sub-processor.[55]

With respect to personal data placed on a cloud service by a customer, cloud providers are normally viewed as processors or sub-processors, processing data on behalf of the customer who acts as controller. This is likely to be more obviously the case further down the cloud supply chain, with PaaS and IaaS providers exercising less control over the customer's processing activity than a SaaS provider. The latter might, depending on the nature of the processing activity and relationship with the customer, be considered a joint controller in respect of certain personal data.[56] Conversely, for personal data generated through use of the cloud service by the cloud customer and its users, as meta-data, the cloud provider will generally act as the controller. Any determination of the regulatory status of a cloud provider will ultimately be based on their actual role in the particular circumstances, rather than any designation contained in contractual statements.[57]

---

[53] See further Michels JD and Walden I (2021) n 46, pp. 396-398.

[54] Art. 5(1)(f) GDPR.

[55] See further Kamarinou D, Millard C, and Turton F 'Responsibilities of Controllers and Processors of Personal Data in Clouds' in: Millard C (ed.) *Cloud Computing Law* (2021, 2nd edn., Oxford University Press).

The GDPR operates on the presumption that the controller issues instructions to the processor. In the cloud market, the service provider as processor will generally determine the conditions under which the service is used and thereby the ability of the controller to meet its compliance obligations. Indeed, given the size of cloud providers such as AWS and Google, GDPR 'compliance-as-a-service' offerings have emerged, whereby service providers provide customers with the necessary assurances for the customer as controller to meet their compliance obligations, including evidencing accountability in terms of appropriate security.[58]

Finally, the GDPR does not apply to companies that merely develop hardware and software for processing personal data, but do not process personal data themselves. Instead, "producers of […] products, services, and applications" are merely "encouraged to take into account the right to data protection".[59] As a result, software developers who offer software 'as a product' are covered neither by the NISD, nor by the GDPR. In contrast, a company that offers the same software 'as a service' would be covered as a cloud provider under the NISD and a processor under the GDPR. The omission of hardware and software developers from the scope of the EU cybersecurity regulatory framework leaves a hole that product liability rules alone are unfit to fill.[60]

---

[56] See C-40/17, Fashion ID GmbH & Co KG v Verbraucherzentrale NRW eV [2020] 1 C.M.L.R. 16; Millard C et al. 'At this rate, everyone will be a [joint] controller of personal data!', (2019) International Data Privacy Law, Vol. 9, No. 4.

[57] EDPB, Guidelines 07/2020 on the concepts of controller and processor in the GDPR, at para. 12.

[58] See Kamarinou D, Millard C, and Oldani, I 'Compliance as a Service', Queen Mary School of Law Legal Studies Research Paper No. 287/2018 (2018).

## 3. SUBSTANCE: CYBERSECURITY OBLIGATIONS FOR CLOUD PROVIDERS

The EU cybersecurity regulatory framework imposes two categories of obligation:

1. safeguarding obligations; and,

2. incident notification obligations.

Broadly speaking, the first category requires cloud providers to take measures to prevent, and mitigate damages from, security incidents. The second is triggered after a provider has suffered a security incident and imposes obligations to notify certain parties. Below, we address each component in turn.

### 3.1 SAFEGUARDING OBLIGATIONS

Regulations typically oblige entities to put in place *appropriate* and *proportionate* security measures. For example,

---

[59] Recital 78 GDPR.

[60] See Fuster GG and Jasmontaite L 'Cybersecurity Regulation in the European Union: The Digital, the Critical and Fundamental Rights' in: Christen M, Gordijn B, and Loi M (eds), *The Ethics of Cybersecurity* (2020, Springer), pp. 109-110; Wolters P 'The security of personal data under the GDPR: a harmonized duty or a shared responsibility?', (2017) International Data Privacy Law, 7:3, pp. 171-172.

under the NISD, DSPs must take "appropriate and proportionate technical and organisational measures" to manage the risks posed to the security of the IT systems they use and ensure a level of security appropriate to the risk posed.[61] Under the GDPR, controllers are responsible for "appropriate security" as a principle of personal data processing;[62] while controllers and processors are required to "implement appropriate technical and organisational measures to ensure a level of security appropriate to the risks" posed to the rights and freedoms of natural persons.[63] On top of this overarching, principles-based safeguarding obligation, the NISD layers more specific security objectives addressed to cloud providers. DSPs must aim for "ensuring the continuity of their services".[64] The Commission's Implementing Regulation further lists required "security elements", including physical security, security of critical supplies, and controls on physical and logical access.[65] The GDPR also spells out specific security objectives for controllers and processors, to ensure the confidentiality, integrity, and availability (or 'CIA') of processing systems and the ability to restore access to personal data.[66] It further mentions pseudonymisation and encryption as measures for protecting personal data, as well as a

---

[61] Art. 16(1)-(2) NISD. The draft NIS 2 Proposal maintains the same safeguarding obligation, at Art. 18(1).

[62] Art. 5(1)(f) GDPR.

[63] Art. 32 GDPR.

[64] Art. 16(1) NISD.

[65] Commission Implementing Regulation (EU) 2018/151 laying down rules for application of Directive (EU) 2016/1148 as regards further specification of the elements to be taken into account by digital service providers, OJ L 26, Art. 2 (hereafter 'Commission NIS Implementing Regulation'). The draft NIS2 Proposal lists relevant "cybersecurity risk management measures" for essential and important entities, which are similar to these "security elements" for DSPs. See Art. 18(2) NIS 2 Proposal.

[66] Art. 32(1)(b) and (c) GDPR.

process for regularly testing and evaluating security measures.[67]

These obligations do not require cloud providers to put in place specific security measures (such as cryptography or two-factor-authentication for access control) or assurance methods (such as external on-site audits or penetration testing). Instead, they require cloud providers to engage in risk management: the ongoing process of identifying, assessing, and responding to risk.[68] As the European Commission put it, the NISD "promotes a risk management culture".[69] The regulations *do* guide cloud providers' risk management by listing factors they should consider. For example, both the NISD and the GDPR require the provider to consider the 'state of the art' and the cost of implementation.[70] Thus, cloud providers must keep abreast of technological developments, but need only adopt cost-effective measures.[71] In effect, safeguarding obligations are more about process than about outcomes. In other words, they are "an obligation of means".[72] To comply, a provider must follow a defined risk management process, analyse relevant information, and make a good-faith determination on how to respond.[73] As Banasiński and Rojszczak put it: because "information security is a process rather than a state", the prevailing regulatory approach is "based on assessing the maturity of the process of ensuring security".[74] To prove compliance, an entity must document its risk management carefully and comprehensively. For example, under NISD, DSPs must have adequate documentation to enable the competent authority to verify compliance with security requirements.[75] Under the GDPR, being able to demonstrate compliance is part of the accountability principle.[76]

---

[67] Art. 32(1) GDPR.

[68] See National Institute of Standards and Technology, 'NIST Framework for improving critical infrastructure cybersecurity v. 1.1', (2018), p. 4.

[69] European Commission, 'Report assessing the consistency of the approaches taken by Member States in the identification of operators of essential services in accordance with Art. 23(1) of Directive 2016/1148/EU on security of network and information systems', (2019) COM(2019) 546 final.

[70] Art. 32(1) GDPR; Art. 16(1) NISD. The Art. 16(1) NISD reference to 'proportionate' measures suggests the cost of measures is a relevant factor, particularly since the Directive aims to avoid imposing disproportionate financial or administrative burdens on the regulated entities. See further NIS Cooperation Group, 'Reference Document on security measures for Operators of Essential Services', CG Publication 01/2018, p. 9.

[71] Kamarinou D et al. (2021) n 55, p. 318.

Such open, principles-based safeguarding obligations can lead to legal uncertainty, since different courts and regulators might interpret them differently.[77] Regulators and other official bodies can reduce this uncertainty somewhat by issuing guidance on how to interpret and implement risk management in practice. For example, in 2016, ENISA published technical guidelines for DSPs under NISD. The guidelines define "common baseline security objectives" and describe measures that DSPs can implement to meet these objectives, with reference to common industry standards such as by the International Organisation for

---

[72] Wolters P (2017) n 60, p. 171.

[73] Michels JD and Walden I (2021) n 46, pp. 409-412.

[74] Banasiński C and Rojszczak M, 'Cybersecurity of consumer products against the background of the EU model of cyberspace protection', (2021) Journal of Cybersecurity, p. 10.

[75] Art. 2(6) Commission NIS Implementing Regulation.

[76] Art. 5(2) GDPR.

[77] Wolters P (2017) n 60, p. 175; Calliess C and Baumgarten A, 'Cybersecurity in the EU The Example of the Financial Sector: A Legal Perspective', (2020) German Law Journal, pp. 1167-1168.

Standardisation ('ISO').[78] National regulators also publish guidance. For example, the Commission Nationale Informatique et Libertés ('CNIL') (the French supervisory authority), has published a guide on securing personal data under the GDPR.[79] Such guidance can be helpful, although it is generally non-binding. However, if different national regulators publish different guidance under GDPR, this can complicate compliance for cloud providers, who are often subject to concurrent national jurisdiction, as discussed below in Section 4.1.

In sum, as DSPs under NISD, cloud providers need to manage the risk that a service disruption would pose to broader societal and economic activities. To do so, they must consider, inter alia, the type of service they provide and their market share. As controllers under the GDPR, cloud providers need to manage the risks to the rights and freedoms of data subjects, taking into account the type of data and the "nature, scope, context and purposes" of their processing activities.[80] What about cloud providers acting as processors? The GDPR imposes an independent obligation to ensure appropriate security on processors, which did not exist under the previous Data Protection Directive.[81] Yet there are limits to how cloud providers acting as processors can exercise the required risk management, in two respects: (i) control and (ii) knowledge. First, since cloud resources are often self-service, a cloud provider necessarily shares control over security aspects with its customers.[82] For example, while an IaaS provider controls security at the physical infrastructure and virtualisation layers, customers control the security of the software they install on virtual machines and configuration settings.[83] As a result, the cloud provider-as-processor's responsibility should be limited to security aspects within its control.

---

[78] ENISA, 'Technical Guidelines for the implementation of minimum security measures for Digital Service Providers', (2016), p.11. See also, ENISA, 'Guidelines on assessing DSP and OES compliance to the NISD security requirements', (2018).

[79] CNIL, 'Security of Personal Data', (2018), https://www.cnil.fr/sites/default/files/atoms/files/cnil_guide_securite_personnelle_gb_web.pdf.

[80] Art. 32(1) GDPR.

[81] Directive 95/46/EC of the European Parliament and of the Council of 24 October 1995 on the protection of individuals with regard to the processing of personal data and on the free movement of such data *OJ L 281, 23.11.1995*.

Second, a cloud provider might have limited knowledge of how its service is being used. Admittedly, in some cases, the nature of the service might indicate its use, such as with a SaaS-service for HR or payroll functions. Yet in other cases, the cloud provider cannot reasonably foresee customer use. For example, imagine IaaS provider 'A' with thousands of customers. A has no idea how each customer is using the self-service resources it makes available. Further, in case of 'cloud layering', provider 'B' might build a SaaS-service on top of A's IaaS service, which B's customer ('C') uses to process personal data. In that case, A would have no contact with C, nor insight into C's processing. What's more, if either B or C encrypts data, A might technically be unable to determine what data is being stored, even if it tried.[84] So, how can a provider as (sub-)processor consider the "nature, scope, context and purposes" of processing? For A to engage in meaningful risk management, each customer would need to inform A of relevant factors such as the types of personal data they are processing. What's more, customers would need to update this information on an ongoing basis, as their cloud use develops. Finally, in case of layered cloud, this information would need to be passed down the supply chain to sub-processors. This is difficult to align with the passive provision of self-service resources on a one-to-many basis as practised by major public cloud providers.[85]

---

[82] ENISA, 'Cloud Security for Healthcare Services', January 2021, p.14.

[83] Hon KW, Millard C, and Singh J, 'Control, Security, and Risk in the Cloud', in: Millard C (ed.) *Cloud Computing Law* (2021, 2nd edn., Oxford University Press), pp. 28-29.

[84] Kamarinou D et al. (2021) n 55, p. 317.

The GDPR tries to achieve such supply-chain transparency by requiring the controller to enter into a contract with the processor that details the "subject-matter and duration" and "nature and purpose" of processing, as well as "the type of personal data and categories of data subjects".[86] The processor is then obliged to pass the same information onto any sub-processor used to carry out processing on behalf of the controller.[87] However, in practice, cloud contracts are typically based on the provider's standard terms. These often describe types of data at a general level, by reference to whatever the customer chooses to upload to the cloud.[88] This generic description provides little input for a provider's risk assessment. This mismatch between processor risk management obligations, on the one hand, and a cloud provider's control and knowledge, on the other, can lead to significant legal uncertainty. In theory, a provider could take a cautious approach and apply the strictest security measures to all its services. However, this would drive up costs of cloud services for all customers, regardless of the risks posed by the relevant processing. That would not be in line with regulations' requirement of a risk-appropriate level of security. Instead, the cloud industry has responded by addressing this uncertainty through GDPR codes of conduct, which we turn to next.

---

[85] Hon KW et al. (2021) n 83, p. 33.

[86] Art. 28(3) GDPR.

[87] Art. 28(4) GDPR.

[88] Hon KW, Millard C, Walden I, Ward C, 'Negotiated Contracts for Cloud Services', in: Millard C (ed.) *Cloud Computing Law* (2021, 2nd edn., Oxford University Press), pp. 116-117.

## 3.1.1 ASSURANCE METHODS: CERTIFICATION AND CODES OF CONDUCT

Assurance refers to a set of actions that bring some level of confidence that certain security requirements are met.[89] A cloud provider can provide assurance that it is fulfilling its safeguarding obligations through certification and/or adherence to a code of conduct. In 2021, ENISA proposed a draft certification scheme for cloud providers under the Cybersecurity Act,[90] and the EDPB issued positive opinions on two codes of conduct under the GDPR.[91] These assurance schemes are voluntary.[92] The GDPR explicitly states that controllers and processors can use adherence to an approved code of conduct to demonstrate compliance with security obligations.[93] The NISD does not explicitly mention certification. Nonetheless, certification under ENISA's scheme should help cloud providers demonstrate compliance with NISD. The draft NIS2 goes one step further: Member States and the Commission can *require* essential entities (which would include cloud providers) to certify ICT services under specific certification schemes to demonstrate compliance.[94] Cloud providers could, in future, face mandatory certification requirements under NIS2. Below, we first review ENISA's draft certification scheme for cloud providers, before reviewing sector-specific codes of conduct for GDPR.

---

[89] ENISA, 'Cybersecurity certification is a global trade and trust instrument', webinar slides, 11 January 2021, https://www.enisa.europa.eu/events/eventfiles/enisa-cybersecurity-certification-of-cloud-services-presentation p. 16.

[90] ENISA, 'EUCS – Cloud Service Scheme', (2020), https://www.enisa.europa.eu/publications/eucs-cloud-service-scheme ('ENISA EUCS').

[91] Opinion 17/2021 on the draft decision of the French Supervisory Authority regarding the European code of conduct submitted by the Cloud Infrastructure Service Providers (CISPE), adopted 19 May 2021 and Opinion 16/2021 on the draft decision of the Belgian Supervisory Authority regarding the "EU Data Protection Code of Conduct for Cloud Service Providers" submitted by Scope Europe. Both adopted on 19 May 2021.

[92] Art. 56(2), Recitals 91-92 Cybersecurity Act.

[93] Art. 32(3), 40, 42 GDPR. The GDPR also mentions "approved certification mechanisms". The authors were not aware of any such mechanisms being approved at time of writing.

The Cybersecurity Act tasks ENISA with setting up a certification framework for ICT services.[95] In December 2020, at the Commission's request, ENISA published a draft certification scheme for cloud services ('EUCS').[96] The EUCS builds on existing national schemes, such as C5 in Germany and SecNumCloud in France.[97] Certification reflects that the cloud provider has met a number of listed security objectives to a certain level of assurance. There are three assurance levels. A *basic* level of assurance can be achieved through the assessment of technical documentation. A *substantial* level requires an additional review of known vulnerabilities and testing through on-site audits. A *high* level of assurance requires an additional assessment of the system's resistance to skilled attackers, by means of penetration testing.[98] The EUCS contains a detailed 80-page list of security measures for each assurance level, such as two-factor authentication ('2FA') for access to non-public areas of physical sites; signature-based and behaviour-based malware protection tools;

---

[94] Art. 21 NIS2 Proposal.

[95] Art. 8 Cybersecurity Act.

[96] European Commission, 'Towards a more secure and trusted cloud in Europe', 9 December 2019, https://ec.europa.eu/digital-single-market/en/news/towards-more-secure-and-trusted-cloud-europe; ENISA EUCS, pp. 49-50.

[97] ENISA, 'Cybersecurity certification is a global trade and trust instrument', webinar slides, 11 January 2021, https://www.enisa.europa.eu/events/eventfiles/enisa-cybersecurity-certification-of-cloud-services-presentation pp. 24, 34-36.

[98] Art. 52(1), (6)-(7), 53(1) Cybersecurity Act; ENISA EUCS, pp. 19-20, 24-26.

data backup at a remote location; and regular penetration tests. The EUCS further includes measures for identity and access management ('IAM'), such as account-specific privileges and encrypted storage of passwords, and for encryption and key management.[99] Despite this level of detail, the EUCS aims to be technology-neutral.[100] This leaves scope for provider discretion in implementation. For example, while the EUCS refers to 2FA, it doesn't require specific techniques (such as strong passwords). That leaves cloud providers free to experiment with effective alternatives, such as 'password-less' authentication based on biometrics. ENISA will publish further guidance on accepted techniques for fulfilling EUCS objectives.[101]

Further, as a sector-specific scheme, the EUCS takes into account the cloud-specific issues we identified above, such as shared responsibility and cloud layering. For example, it requires providers to explicitly and publicly document the allocation of responsibility for security between provider and customer.[102] Further, SaaS-providers who rely on an underlying IaaS-provider face a simplified certification process if they use a certified sub-provider. Provided the split of responsibilities between the providers is defined, the SaaS-provider can rely on the underlying sub-provider's certification for relevant aspects of the system (such as access controls on physical sites).[103]

Under the GDPR, national supervisory authorities can approve codes of conduct for processing activities in their territory. For processing activities in multiple Member

---

[99] ENISA EUCS, pp. 81-159.

[100] ENISA EUCS, p. 82.

[101] ENISA EUCS, p. 82.

[102] ENISA EUCS, p. 9.

[103] ENISA refers to this as 'composition' on a 'base cloud service', see ENISA EUCS, pp.74-75.

States, the national supervisory authority shall first seek an opinion from the European Data Protection Board ('EPDB').[104] The Commission can then decide that the approved code of conduct is valid across the EU.[105] In May 2021, the EDPB adopted favourable opinions on two codes of conduct for cloud services.[106] In June 2021, the CNIL approved the Cloud Infrastructure Service Providers in Europe Code ('CISPE').[107] CISPE is tailored specifically to IaaS providers acting as processors.[108] Its members include AWS and some 25 smaller providers, such as OVH.[109] Second, in May 2021, the Belgian Data Protection Authority ('SA') approved the EU Cloud Code of Conduct ('EUCoC').[110] The EUCoC covers all B2B cloud service types (i.e. IaaS, PaaS, and SaaS) where the provider acts as a processor.[111] Its members include Microsoft and Google Cloud, as well as Alibaba Cloud, Cisco, Dropbox, IBM, Oracle, Salesforce, and SAP.[112] Like the EUCS, the CISPE and EUCoC list security objectives that cover a range of topics including physical security, logical access management, encryption, vulnerability management, and logging and monitoring. The codes also provide examples of relevant security measures, with reference to international standards such as ISO.[113]

---

[104] Art. 40(6)-(7) GDPR.

[105] Art. 40(9) GDPR.

[106] EDPB, 'EDPB adopts opinions on first transnational codes of conduct', 20 May 2021, https://edpb.europa.eu/news/news/2021/edpb-adopts-opinions-first-transnational-codes-conduct-statement-data-governance-act_en.

[107] CNIL, 'The CNIL approves the first European code of conduct for cloud infrastructure service providers (IaaS)' (2021), https://www.cnil.fr/en/cnil-approves-first-european-code-conduct-cloud-infrastructure-service-providers-iaas .

[108] CISPE Code, 9 February 2021, available at https://cispe.cloud/, pp. 5-6.

[109] https://cispe.cloud/members/.

[110] Belgian Data Protection Authority, 'Approval decision of the "EU Data Protection Code of Conduct for Cloud Service Providers"' (2021), Decision n° 05/2021 of 20 May 2021 https://www.dataprotectionauthority.be/publications/decision-n05-2021-of-20-may-2021.pdf.

[111] EUCoC, pp. 3-4.

[112] https://eucoc.cloud/en/home/.

These sector-specific codes also take into account cloud-specific issues. For example, the CISPE Code sets out a shared responsibility model for IaaS-providers that addresses the complications of knowledge and control we described above. Specifically, CISPE recognises that IaaS-providers "have no control" over whether customers use the service for processing personal data and, if so, for what purpose.[114] Only the customer knows its intended processing in sufficient detail to determine an appropriate level of security.[115] To address this, the CISPE Code allocates responsibilities between the IaaS provider and the customer. Broadly speaking, the provider is responsible for the physical infrastructure, the virtualisation layer, and any host operating software. We refer to this as the 'host system'. Conversely, the customer is responsible for software it installs on the host system.[116] We refer to this as 'guest software'. This distinction between responsibility for the host system and guest software resembles AWS's distinction between the provider's responsibility for security '*of* the cloud' and the customer's responsibility for security '*in* the cloud'.[117] Under the Code, the provider must implement a "baseline" of security measures for the host system and can offer additional "enhanced" measures for customers to select and apply.[118] As under the EUCS, the provider must clearly document its measures and the split of responsibilities. The customer has ultimate responsibility for ensuring that an appropriate overall level of security is in place for its processing activities.[119]

---

[113] CISPE Code, Annex A, pp. 58-71; EUCoC, pp. 17-20, 24, Annex A, pp. 29-46.

[114] CISPE Code, p. 14.

[115] CISPE Code, p.19.

[116] CISPE Code, p.18.

[117] AWS, 'Shared Responsibility Model', https://aws.amazon.com/compliance/shared-responsibility-model/.

[118] CISPE Code, p.18.

The EUCoC similarly implies a shared responsibility model. As under CISPE, the provider must provide a "baseline" of security and must transparently describe the security measures in place. The provider can offer further "security options", which customers can choose to apply.[120] The EUCoC further acknowledges the complications IaaS-providers face when acting as processors, by stating that providers must take account of their "knowledge (if any) of the sensitivity of the Customer Personal Data being processed", including by considering "the nature of the Cloud Service" and the impact of a data breach "insofar as this is known to the [provider]". By way of example, a provider might gain actual knowledge of sensitive personal data being processed through negotiations with the customer.[121] This implies that, in other situations, the provider might not have relevant knowledge at all.

In sum, the EUCS and GDPR Codes of Conduct provide welcome legal certainty in three respects. First, they provide a greater level of *specificity* as to the objectives and kinds of measures that constitute 'appropriate and proportionate' security for cloud providers. This provides concrete guidelines as to how cloud providers can comply with the over-arching principles-based safeguarding obligations under NISD and GDPR. Second, the schemes provide a greater degree of *uniformity*, by applying a single set of guidelines across the EU. This matters since cloud providers can find themselves subject to multiple concurrent national jurisdictions, as discussed below in Section 4.1. Third, the schemes take into account *cloud-specific factors* that limit a provider's ability to manage risk, including a lack of knowledge and control. In particular, each scheme supports a shared responsibility model, that differentiates between the provider's host system and the customer's guest software and data. Under GDPR, the customer-as-controller remains responsible for deciding whether (i) the provider's baseline security is sufficient for its processing, (ii) whether to use the provider's enhanced security options; or (iii) whether to avoid use the cloud service at all. In other words, each scheme aims to support the customer's assessment of whether the provider's security measures meet the customer's security requirements.[122]

---

[119] CISPE Code, p.9
[120] EUCoC, pp. 4, 9, 17, 20.
[121] EUCoC, p. 17.

The approved Codes of Conduct have significant uptake, including from all three major IaaS providers. It remains to be seen whether the draft EUCS would see such widespread adoption. However, over time, EUCS certification could evolve into a *de facto* standard. For instance, customers in regulated sectors like financial services or OESs might require certification to a high level as a prerequisite for their use of cloud services. Further, the EUCS might not, in practice, require major cloud providers to change much. Such providers typically already submit themselves to annual external audits and make audit reports (such as SOC 1 and 2) available to (potential) customers.[123] One main difference between these current practices and the EUCS would be that the audit will cover security objectives and controls - and will be done to a standard - publicly set by ENISA, instead of by the cloud provider. That might promote trust in the cloud. Admittedly, obtaining EUCS certification might be more challenging for small cloud providers. That said, the EUCS could also simplify their risk management. Instead of weighing up 'appropriate and proportionate' responses to a range of possible risks, they need only determine whether their service presents a low, medium, or high risk and obtain certification accordingly.[124] The EUCS then provides ready-made security objectives and controls, while ENISA recommends accepted techniques.

---

[122] ENISA EUCS, p. 9; CISPE Code, p. 9; EUCoC, p. 2.
[123] Hon KW et al. (2021) n 88, pp. 127-132.

## 3.2 INCIDENT NOTIFICATION

Both the NISD and the GDPR require entities to notify cybersecurity incidents. An incident refers to any set of circumstances that compromises cybersecurity.[125] The requirements differ in relation to: (i) who the entity must notify and (ii) what kind of incident triggers a notification obligation, which we refer to as a 'reporting threshold'. Under NISD, DSPs must notify the competent authority or the national Computer Security Incident Response Team ('CSIRT') of security incidents that have a substantial impact on the provision of its service.[126] Whether an impact is *substantial* depends on factors like the number of affected users, the incident's duration and geographic spread, and the extent of the service disruption and its impact on economic and societal activities.[127] The Commission has specified that DSPs must report an incident if: (i) their service is unavailable for more than 5 million user-hours; or the incident (ii) affected the CIA of data relating to more than 100,000 users; (iii) created a risk to public safety, security, or loss of life; or (iv) caused material damage of at least €1m.[128] In some cases, cloud providers can face difficulties in assessing the impact of a service disruption. As noted above, an IaaS provider might have no idea how customers are using its service.[129] The NISD recognizes this: incident notification requirements only apply when DSPs have access to the information needed to assess an incident's impact.[130] DSPs are not required to collect additional information to do so, but should consider their contractual relationships with customers and their traffic data.[131] As a result, IaaS providers with limited knowledge of how their customers use their service could argue that they are exempt from notifying security incidents under the NISD. This presents a risk of under-reporting, as regulators miss information on security incidents that impact IaaS providers. This risk is particularly problematic considering the market concentration at the IaaS-layer, as set out above in Section 2.1. Further, under the GDPR, a controller must notify the relevant supervisory authority of a personal data breach that is likely to result in a risk to the rights and freedoms of natural persons.[132] In addition, the controller must notify the affected data subjects in case a personal data breach is likely to result in a *high* risk to the rights and freedoms of natural persons.[133] Conversely, a processor must notify a controller after becoming aware of any personal data breach (regardless of any risks posed).[134]

---

[124] Art. 52(1) Cybersecurity Act: "The assurance level shall be commensurate with the level of the risk associated with the intended use of the ICT service".

[125] Art. 4(7) NISD.

[126] Art. 16(3) NISD.

[127] Art. 16(4) NISD.

[128] Commission NIS Implementing Regulation, Art. 4.

The NISD and GDPR therefore present three different incident reporting thresholds. First, the NISD features a quantifiable, outcome-based threshold: substantial impact. Second, the GDPR obligation for controllers has a risk-based threshold: a (high) risk to natural persons. Third, the GDPR obligation for processors applies to every incident that affects personal data, without any further requirements.[135] This means that incidents that do not substantially disrupt a cloud provider's service would not trigger NISD's notification. However, such non-disruptive incidents could still reveal serious shortcomings in the provider's security. For example, incidents of unauthorised access might impact the confidentiality of highly sensitive data of a small number of customers, without leading to wider service disruption. Further, a cloud provider might suffer a ransomware attack, but pay the ransom before it caused any disruption to its service. In such cases, regulators risk missing information about incidents that point to serious security failures.[136] Conversely, under the GDPR, a cloud provider acting as processor would need to notify a customer acting as controller of any such incidents that involved personal data. The customer then assesses the risk posed to natural persons and decides whether to notify regulators and data subjects, as appropriate.[137] If the non-disruptive incident posed a risk to natural persons, for instance because of a confidentiality breach, the controller would need to notify the regulator. If not, the controller would at least need to consider the non-disruptive incident as part of its risk management process, including in determining whether the provider's security measures meet its requirements.

---

[129] Michels JD and Walden I (2021) n 46, pp. 397-399.

[130] Art. 16(4) NISD.

[131] Commission NIS Implementing Regulation, Art. 3(6).

[132] Art. 33(1) GDPR.

[133] Art. 34(1) GDPR.

[134] Art. 33(2) GDPR.

The draft NIS2 proposes two changes to incident notification thresholds. First, it re-defines a *significant* impact as including an incident that "has the potential to cause" disruption.[138] This shift to a risk-based threshold would address the two risks of under-reporting identified above, namely in relation to IaaS providers without access to information about concrete impacts and in relation to non-disruptive incidents. Second, it proposes a *new* notification obligation for entities to notify "any significant cyber threat" that "could have potentially resulted in a significant incident".[139] In this context, a *cyber threat* means any circumstance, event or action that could adversely impact IT systems, users, or other persons.[140] The rapporteur has criticized these proposals and called for their removal. The rapporteur argues that reporting incidents which only have a 'potential impact' could lead to "[o]ver-reporting of potential incidents that have not happened", which could overwhelm regulators.[141] We disagree. In our view, the revised obligation would only apply to *actual* security incidents (that is: breaches of CIA) of a serious nature, which could have caused widespread disruption, but did not do so for reasons other than the provider's security measures – as in the examples above. This can help alert regulators to serious security failures. Nonetheless, we agree that reporting mere 'threats' as opposed to actual 'incidents' is overly broad, since cloud providers presumably face a wide range of threats on a daily basis. Reporting mere 'threats' could therefore lead to regulators being overwhelmed by notification overload.[142]

---

[135] Art. 4(12) GDPR.

[136] Michels JD and Walden I (2021) n 46, pp. 412-414.

[137] Kamarinou D et al. (2021) n 55, p. 323.

[138] Art. 20(1) and (3)(a) NIS 2 Proposal.

---

[139] Art. 20(2) NIS 2 Proposal.

[140] Art. 4(7) NIS 2 Proposal.

[141] Groothuis B, 'Draft report on the proposal for a directive of the European Parliament and of the Council on measures for a high common level of cybersecurity across the Union', 3 May 2021, 2020/0359(COD), pp. 38-39 ('Rapporteur Groothuis Report on NIS2 Proposal').

[142] Rapporteur Groothuis Report on NIS2 Proposal, pp. 37-38.

# 4. ENFORCEMENT: OVERSEEING CLOUD PROVIDERS' COMPLIANCE

The institutional framework of EU cybersecurity oversight and enforcement is complicated. It involves multiple bodies across the Member State and EU level and can differ per sector. In addition, cloud providers who offer services in multiple Member States can be subject to the concurrent jurisdiction of several national regulators. In this section, we first look at jurisdiction, before turning to oversight and remedies for non-compliance.

## 4.1 JURISDICTION

Under the NISD, a DSP is supervised by a single competent authority in the EU. A cloud provider falls under the jurisdiction of the Member State in which it has its *main establishment*, typically meaning its head office.[143] Consider a major cloud provider who offers services to customers across the EU, operates data centres in various Member States, and has its European headquarters in Dublin. It would fall under the jurisdiction of the Irish regulator: the National Cyber Security Centre ('NCSC'), which is part of the Irish Government Department of Communications.[144] A provider which is not established in the EU, but offers cloud services in the EU, must designate an EU representative in a Member State. It then falls under that Member State's jurisdiction.[145] As a result, cloud providers are subject to the exclusive jurisdiction of one NISD regulator. The draft NIS2 maintains this system.[146]

---

[143] Recital 64 NISD provides that a DSP's main establishment "in principle corresponds to the place where the provider has its head office in the Union".

[144] European Commission, 'Implementation of the NIS Directive in Ireland', https://digital-strategy.ec.europa.eu/en/policies/nis-directive-ireland.

Jurisdiction is more complicated under the GDPR. In principle, each national supervisory authority is competent for processing activities that affect data subjects on its territory.[147] However, the GDPR establishes a 'one-stop-shop' ('OSS') mechanism for cross-border processing: the supervisory authority of the Member State where a cloud provider has its *main establishment* acts as the provider's *lead supervisory authority*.[148] In the above example of a major cloud provider headquartered Dublin, the lead authority is the Irish supervisory authority: the Data Protection Commission ('DPC').[149] The lead authority has primary responsibility for coordinating investigations and decisions on enforcement, while cooperating with "concerned" supervisory authorities in other Member States. The GDPR sets out a cooperation procedure, which involves sharing information and discussing draft decisions to try to reach consensus.[150]

However, the lead supervisory authority does not have exclusive jurisdiction for cloud providers, in four respects. First, the EDPB can adopt a binding decision that overrides a decision of the lead supervisory authority,[151] which it has done in relation to a decision of the DPC concerning WhatsApp Ireland.[152] Second, other supervisory authorities retain a residual competence to handle complaints from data subjects, if the subject matter substantially affects data subjects in one Member State only.[153] For example, a cloud provider headquartered in Dublin could experience a local cloud outage that only affects data subjects in France. The CNIL is then competent to consider related complaints from data subjects, although the DPC is the lead authority. Third, as noted above, a cloud provider will typically process personal data as a processor on behalf of customers who act as controllers and who are based in multiple Member States. Under Recital 36, "in cases involving both controller and processor", the lead supervisory authority is the *controller's* lead authority (*not* the processor's), which effectively means that the processor might have to deal with multiple supervisory authorities.[154] Since cybersecurity is a shared responsibility between a cloud provider and its customers, an incident "involving both the controller and the processor"[155] is likely to be a common occurrence. As a result, cloud providers acting as processors can be subject to the jurisdiction of their customers' lead authorities. Fourth, cloud providers who are not established in the EU cannot benefit from the one-stop-shop mechanism; they are supervised by local supervisory authorities in every Member State.[156] It is less clear what happens when a cloud provider *is* established in a Member State, but acts as a processor for a controller who *is not* established in a Member State. Following the logic that (i) the non-EU controller does not benefit from the one-stop-shop mechanism and (ii) in cases involving both controller and processor, the processor is subject to the controller's competent authority, it would appear that (iii) the cloud provider is then also subject to the jurisdiction of supervisory authorities in every Member State.

---

[145] Art. 18 NISD. Recital 65 clarifies that the mere accessibility of services in the EU is insufficient to determine whether a DSP is offering services in the EU, but that other factors, such as currency and language of the service or the mentioning of customers in the EU, are relevant.

[146] Art. 24(1)-(3) NIS 2 Proposal, pp. 10-11.

[147] Recital 122 GDPR.

[148] Art. 56 GDPR.

[149] Data Protection Commission, https://www.dataprotection.ie/.

[150] Art. 60 GDPR.

[151] Art. 65(1)(a) GDPR.

---

[152] EDPB, Binding decision 1/2021 on the dispute arisen on the draft decision of the Irish Supervisory Authority regarding WhatsApp Ireland under Article 65(1)(a) GDPR (28 July 2021). See also Decision 1/2020 (9 November 2020) concerning a decision by the DPC regarding Twitter International, in which the DPC's decision was upheld.

[153] Art. 56(2) and (3) GDPR.

[154] Article 29 Data Protection Working Party, 'Guidelines for identifying a controller or processor's lead supervisory authority', WP244 rev.01 (2017), adopted by the EDPB on 25 May 2018, https://edpb.europa.eu/our-work-tools/our-documents/guidelines/guidelines-identifying-controller-or-processors-lead_en, p.9 ('A29WP Guidelines for identifying lead supervisory authority').

[155] Recital 36 GDPR.

In sum, jurisdiction under both NISD and GDPR is based on the cloud provider's main establishment. However, while the NISD regulator's jurisdiction is exclusive, the GDPR lead authority only has primary responsibility. This effectively renders the GDPR's OSS mechanism meaningless for cloud providers in their role as processors, given the self-service and one-to-many nature of cloud services. Since a cloud provider might not know where its customers' main establishments are, it is difficult to predict which supervisory authorities will exercise jurisdiction over an incident . That can, in turn, complicate compliance with safeguarding obligations, particularly if different supervisory authorities each publish their own cybersecurity guidelines.

The draft EUCS and approved Codes of Conduct might help address this uncertainty, by providing uniform security requirements applicable across the EU.[157] While potentially simplifying compliance, the schemes add another layer of institutional complexity. For example, under the draft EUCS, three types of bodies are involved in issuing certificates and monitoring compliance: conformance assessment bodies ('CABs'), national accreditation bodies ('NABs') and the national cybersecurity certification authority ('NCCA'). Each Member State shall appoint a NAB[158] and an NCCA.[159] NABs accredit and monitor CABs who, in turn, issue certificates to cloud providers.[160] A cloud provider can choose to submit its application for certification to a CAB anywhere in the EU. Once issued, the certificate is valid across the EU.[161] The CAB who issued the certificate and the NCCA shall supervise and enforce compliance with the certification scheme rules and suspend or withdraw certificates where needed.[162] The NCCA shall have the power to request information from and audit certificate holders as required.[163]

---

[156] A29WP Guidelines for identifying lead supervisory authority, p. 10.

[157] See e.g. Recital 95 Cybersecurity Act.

      The GDPR Codes of Conduct also introduce new oversight bodies. Under CISPE and EUCoC, a cloud provider can be confirmed as compliant by a Monitoring Body ('MB').[164] The CNIL accredits MBs under CISPE, while the Belgian SA accredits MBs under the EUCoC.[165] In 2021, the CNIL appointed EY Certifypoint as MB under CISPE, while the Belgian SA appointed Scope Europe as MB under EUCoC.[166] The provider can apply to an MB of their choice.[167] The

---

[158] Regulation (EC) No 765/2008 setting out the requirements for accreditation and market surveillance relating to the marketing of products and repealing Regulation (EEC) No 339/93, L 218/30 13.8.2008, Art. 4.

[159] Art. 58 Cybersecurity Act.

[160] Art. 60, Recital 65 Cybersecurity Act; ENISA EUCS, pp.26, 29.

[161] Recitals 73, 97 Cybersecurity Act.

[162] ENISA EUCS, pp. 39-48.

[163] Art. 58(7)-(8) Cybersecurity Act.

[164] CISPE Code, pp. 40-41; EUCoC, pp. 21-23.

[165] CISPE Code, pp. 47-48; EUCoC, p. 21.

[166] CNIL, 'Code of conduct: CNIL grants first accreditation to a monitoring body', 16 July 2021, https://www.cnil.fr/en/code-conduct-cnil-grants-first-accreditation-monitoring-body; Belgian SA, 'Accreditation of the "Scope Europe" for the monitoring of the "EU Cloud Code of Conduct"', Decision n° 06/2021 of 20 May 2021, https://www.dataprotection-authority.be/publications/decision-n-06-2021-of-20-may-2021.pdf.

[167] CISPE Code, pp. 48, 50.

provider is subject to an initial assessment and then annual review by the MB.[168] In case of non-compliance, the MB can issue warnings or reprimands, suspend the provider's declaration of adherence, or exclude the provider from the Code entirely.[169] In sum, should a cloud provider chose to certify under EUCS and adhere to a GDPR code of conduct, they would become subject to the supervision of three new authorities, namely: a CAB, an NCCA, and an MB.

## 4.2 OVERSIGHT AND ENFORCEMENT

Under the NISD, regulators are expected to take a light-touch, reactive supervisory approach to DSPs. Instead of ongoing supervision, regulators should only act *ex post* when provided with evidence that a DSP is not taking appropriate measures.[170] When presented with such evidence, the regulator can investigate and, in case of non-compliance, issue an administrative fine. The NISD does not specify the level of fines. It states only that Member States must provide for penalties that are effective, proportionate, and dissuasive.[171] This has led to divergent national implementations, with maximum fines under Member State law ranging from €5m to a mere €1,400.[172] The draft NIS2 changes this supervisory and enforcement regime. Cloud providers would instead become subject to ongoing, *ex ante* and *ex post* supervision as essential entities.[173] It further aims to harmonise the level of fines: up to €10m or 2% of annual turnover, whichever is highest.[174]

---

[168] CISPE Code, pp. 42-43; EUCoC, pp. 24-25,27.

[169] CISPE Code, pp. 52-54; EUCoC, pp.28-29.

[170] Art. 17 NISD; Commission, 'Communication – Making the most of NIS' (2017) COM(2017)476, 13.9.2017, Annex 1, p. 35; ENISA, 'Incident notification for DSPs in the context of the NIS Directive' (2017), p. 9.

[171] Art. 21 NISD.

[172] Commission, 'Impact Assessment Report accompanying the document Proposal for a Directive of the European Parliament and of the Council on measures for a high common level of cybersecurity across the Union' Part 1 of 3, 16 December 2020, SWD(2020) 345 final, pp. 25-26 ('Commission Impact Assessment').

The draft NIS2 also introduces new obligations and remedies that target management personnel as *natural persons*, rather than cloud companies as *legal persons*. It provides that management bodies must approve NISD risk management decisions, oversee their implementation, and be accountable for non-compliance. They must also follow regular cybersecurity trainings.[175] Further, the regulator can impose a temporary ban from exercising managerial functions against a CEO or legal representative. In addition, Member States shall ensure that natural persons who represent, exercise control of, or take decisions on behalf of, cloud providers "may be held liable for breach of their duties to ensure compliance" with the Directive.[176] The proposal does not specify the type of liability. This could, in theory, range from civil liability for damages to third parties; or personal liability to shareholders for failing in a fiduciary duty; to regulatory liability in the form of administrative fines; or possibly even criminal liability.[177] This shift to holding individuals responsible is likely to prove highly controversial. The draft rapporteur objected to the remedy of banning management personnel as disproportionate.[178] It remains to be seen how the proposal will develop.

The GDPR also features administrative fines. There are two levels of fines: the lower level of up to €10m or 2% of annual turnover and the higher level of up to €20m or 4% of annual turnover (in both cases, whichever is higher).[179] The applicable level of fines for non-compliance with cybersecurity requirements will depend on the circumstances of the case. The lower level applies to breaches of the safeguarding and notification duties under Articles 32-34; while the higher level applies to breaches of the security principle under Article 5(f). This implies that a cloud provider acting as a processor should only be liable to the lower threshold for failure to fulfil its independent obligation to ensure appropriate security under Article 32, as compliance with the principles under Article 5 is the controller's responsibility. Whether regulators will in practice treat these cybersecurity obligations as distinct when determining penalties remains to be seen.

---

[173] Recital 70, Art. 29 NIS 2 Proposal, p. 11.

[174] Art. 31 NIS 2 Proposal.

[175] Art. 17 NIS 2 Proposal.

[176] Art.29(5)-(6) NIS 2 Proposal.

[177] The Commission did not address this point in detail. See e.g. Commission Impact Assessment Part 1, pp. 64-65.

[178] Rapporteur Groothuis Report on NIS2 Proposal, p. 52.

Finally, the draft EUCS and GDPR Codes of Conduct provide further oversight and enforcement. In both cases, non-compliance with security obligations can lead to withdrawal of certificates, which could lead to significant reputational damage. Further, the NCCA can impose penalties under the draft EUCS.[180] The Cybersecurity Act further states that Member States shall put in place effective, proportionate, and dissuasive penalties for infringements of certification schemes.[181] Thus, in future, a certified cloud provider could be subject to administrative fines for cybersecurity under NISD, GDPR, and EUCS. In theory, these fines could relate to the same incident, although this raises concerns in light of the 'double jeopardy' or *ne bis in idem* principle, i.e. that a person should not be tried twice for the same offence. Cumulative fines might be justified however where the regulations "relate to different aspects of the wrongdoing and different impacts".[182] The draft NIS2 would remove the risk of double fines under NISD and GDPR. Where a supervisory authority imposes a fine under GDPR, the NISD regulator shall not impose a fine for the same infringement.[183]

---

[179] Art. 83(4)-(5) GDPR.
[180] Art. 58(8) Cybersecurity Act.
[181] Art. 65 Cybersecurity Act.

To date, the enforcement of the cybersecurity regulatory framework for cloud providers presents a mixed bag. The Commission has characterized the NISD supervision and enforcement regime as "ineffective".[184] As of December 2020, the Commission was not aware of a single NISD penalty.[185] In contrast, GDPR supervisory authorities *have* issued fines for non-compliance with security obligations.[186] Nonetheless, some commentators criticise perceived under-enforcement of the GDPR by specific lead authorities. For example, the Irish Council for Civil Liberties ('ICCL') argues that the DPC has failed to take sufficient action as the lead authority for cross-border GDPR cases, thereby presenting an enforcement bottleneck. It states that: "enforcement against Big Tech is paralysed by Ireland's failure to deliver draft decisions on cross-border cases."[187]

---

[182] DCMS, 'Security of Network and Information Systems: Government response to the public consultation' (January 2018), p. 16. See also Cole M and Schmitz-Berndt S 'The Interplay between the NIS Directive and the GDPR in a Cybersecurity Threat Landscape', (2019) University of Luxembourg Law Working Paper Series Paper number 2019-017, p. 18.

[183] Art. 32(2) NIS 2 Proposal.

[184] NIS 2 Proposal, p. 5.

[185] Commission Impact Assessment Part 1, p. 26.

[186] CNIL, 'Credential stuffing: la CNIL sanctionne un responsable de traitement et son sous-traitant', 27 January 2021, https://www.cnil.fr/fr/credential-stuffing-la-cnil-sanctionne-un-responsable-de-traitement-et-son-sous-traitant. In the UK, in October 2020, the ICO fined British Airways £20m for data breach affecting more than 400,000 customers (see https://ico.org.uk/about-the-ico/news-and-events/news-and-blogs/2020/10/ico-fines-british-airways-20m-for-data-breach-affecting-more-than-400-000-customers); and Marriott International Inc. £18.4million for failing to keep customers' personal data secure (see https://ico.org.uk/about-the-ico/news-and-events/news-and-blogs/2020/10/ico-fines-marriott-international-inc-184million-for-failing-to-keep-customers-personal-data-secure/).

The problem is not per se that regulators lack powers, but that they are not using them. This might be because regulators lack resources. In a survey of 46 NISD regulators conducted in September 2020, the most cited challenge to implementing NISD was "limited resources (staff and financial)".[188] The draft NIS2's new powers targeting management personnel directly are unlikely to help, if regulators lack the resources to make use of their existing powers. What's more, if regulators are expected to engage in ongoing, proactive oversight of all cloud providers under the draft NIS2, this will place further strain on already limited resources. This raises a question: might a narrower objective of *ex post* enforcement only, backed by sufficient resources and (the threat of) high fines, lead to better outcomes? Alternatively, if regulators must apply both *ex ante* and *ex post* enforcement, might a narrower definition of cloud services be more appropriate, for example by restricting the Directive's scope only to the most critical, hyperscale B2B cloud providers? Unfortunately, the current draft NIS2 risks the worst of both worlds: a broad scope covering all cloud providers, coupled with an unrealistic expectation that regulators will apply proactive oversight to the entire sector.

The UK government has proposed a different approach for amending its NIS Regulations. which implemented the NISD into national law in 2018.[189] Since the UK left the EU in 2020, it is not bound to implement the draft NIS2, but can amend its regulations independently. In January 2022, the UK government proposed applying proactive supervision only to the most critical DSPs, and reactive supervision to others.[190] The ICO would identify the most critical DSPs, using criteria similar to those for designating OES under the current NISD, such as: the number of users relying on the service; market concentration; the level of dependence customers place on the service; the level of connectivity and access to the customer's systems; and the likely consequences for national security of service disruptions. This approach reserves proactive supervision for the most critical cloud providers, thereby focusing regulatory attention on those services that present the biggest risk of societal disruption. The draft NIS2 might lead to similar outcomes in practice, as European regulators focus their limited resources on proactively supervising those DSPs that pose the biggest threat to the interests protected by the NIS Directive.[191]

---

[187] ICCL, 'Europe's enforcement paralysis' (2021), https://www.iccl.ie/wp-content/uploads/2021/09/Europes-enforcement-paralysis-2021-ICCL-report-on-GDPR-enforcement.pdf.

[188] Commission Impact Assessment, Part 3, pp. 5,12.

[189] The Network and Information Systems Regulations 2018, 2018 No. 506.

Hopefully, new oversight bodies under EUCS and the approved Codes of Conduct will provide new resources for oversight and enforcement. Cloud providers will pay CABs and MBs for initial review and ongoing oversight. However, Kamara *et al.* point out that such industry funding can undermine trust in the certification process, which:

> "suffers from a fundamental ambiguity insofar as the applicant is also a client. The certification body is a service provider that must, at the same time, satisfy a paying client and scrutinise its procedures with impartiality. The certification body must thus constantly balance the need to ensure the quality of the certification process and the requirement to satisfy a client that is able to swap, at any time, from one provider to another."[192]

---

[190] DCMS, 'Proposal for legislation to improve the UK's cyber resilience', Consultation, 19 January 2022, https://www.gov.uk/government/consultations/proposal-for-legislation-to-improve-the-uks-cyber-resilience/proposal-for-legislation-to-improve-the-uks-cyber-resilience.

[191] Michels JD and Walden I 'Beyond Complacency and Panic: Will the NIS Directive Improve the Cybersecurity of Critical National Infrastructure?' (2020), 45 E.L. Rev, p. 44.

Nonetheless, given the above enforcement issues, additional oversight is welcome. Further, participation in the draft EUCS and Codes of Conduct merely *demonstrates* compliance, it does not *guarantee* it. Only regulators and courts can determine compliance with NISD and GDPR. In sum, while industry funding provides additional resources for supervision, publicly-funded bodies provide a regulatory backstop.

## 5. CONCLUDING REMARKS

In this chapter, we have examined how three of the main legislative instruments of the European cybersecurity regulatory framework (the NISD, the GDPR, and the Cybersecurity Act) impact the provision of cloud computing services. First, in terms of scope, the NISD adopts a structural approach to cybersecurity regulation, with application of the regime triggered by the role that the cloud provider plays within a supply chain, as an 'enabling' service for its users. In contrast, the GDPR adopts a substantive approach, the regime being triggered by the cloud provider's role in processing personal data. Cloud providers are typically subject to both NISD and GDPR. A key concern with the NISD approach is that the resulting regime is a blunt instrument that imposes asymmetric regulatory burdens on a vast range of market participants (especially the long tail of SaaS service providers) that are no more critical or enabling than other parts of the supply chain, such as outsourcing providers and hardware and software producers. Nonetheless, it is arguably the best available approach, given the complexities of determining whether each individual cloud service is crucial enough for the businesses that rely on it that it should qualify as 'critical' infrastructure. Further, since the NISD imposes risk management obligations, cloud providers need only determine the level of security 'appropriate' to the risk which disruption of their service would pose, which should be lower for providers of less 'critical' cloud services. Admittedly, for those operating at the lower layers of the stack, such as IaaS providers, such risk analysis is inherently speculative, since they might not know for which purposes customers are using the service.

---

[192] Kamara I, Leenes R, Lachaud E, Stuurman K, van Lieshout M, Bodea G, 'Data Protection Certification Mechanisms: Study on Articles 42 and 43 of the Regulation (EU) 2016/679' (2019) Report for European Commission, p. 141.

Second, in terms of substance, while the NISD and GDPR impose similarly-worded obligations, they comprise high level principles that inevitably require elaboration in the form of further regulations, guidance, and codes, as well as assurance mechanisms, such as codes of conduct and certification schemes. Together with a multiplicity of regulatory authorities claiming jurisdiction, both at a domestic level and between Member States, such variety is likely to result in divergent regulatory approaches and unintended outcomes, as well as imposing unnecessary compliance costs on cloud providers. For many years, security professionals bemoaned the lack of attention policy makers and legislators paid to cybersecurity. Unfortunately, the EU's current rush to regulate may prove as much of a curse as a gift for those trying to protect our systems, services, and data.